\documentclass[aip,amsmath,amssymb,preprint]{revtex4-1}

\usepackage{amssymb}
\usepackage{amsmath}
\usepackage{bm,xcolor}
\usepackage{graphicx}% Include figure files

\def\func#1{\mathop{\rm #1}}
\usepackage{amsmath,amsfonts,amssymb,epsfig,subfigure}
\usepackage{graphicx}

\newcommand{\be}{\begin{equation}}
\newcommand{\en}{\end{equation}}

\renewcommand{\vec}[1]{\boldsymbol{#1}}
\def \bm#1{\mbox{\boldmath{$#1$}}}   % this is used to write boldface Greek

\begin{document}

{\centerline{\footnotesize\emph{Physics of Fluids 25, 082003 (2013); doi: 10.1063/1.4818159}}}

\title{\centerline{Scaling Navier-Stokes Equation in Nanotubes}}

\author{Mihail G\u{a}r\u{a}jeu}
\author{Henri Gouin}
\affiliation{Aix-Marseille Universit\'e, CNRS, Centrale Marseille, M2P2 UMR 7340, 13451, Marseille, France}
\author{Giuseppe Saccomandi}
\affiliation{Dipartimento di Ingegneria Industriale,
Universit\`{a} degli Studi di Perugia,
          \\ 06125 Perugia, Italy.}
           \email{mihai.garajeu@univ-amu.fr; henri.gouin@univ-amu.fr; giuseppe.saccomandi@unipg.it}

\begin{abstract}
  On one hand, classical Monte Carlo and molecular dynamics (MD)
  simulations have been very useful in the study of liquids in
  nanotubes, enabling a wide variety of properties to be calculated in
  intuitive agreement with experiments. On the other hand, recent
  studies indicate that the theory of continuum breaks down only at the
  nanometer level; consequently flows through nanotubes still can be
  investigated with Navier-Stokes equations if we take
  suitable boundary conditions into account.
  \newline
  The aim of this paper is to
  study the statics and dynamics of liquids in nanotubes by using
  methods of non-linear continuum mechanics. We assume that the
  nanotube is filled with only a liquid phase; by using a second
  gradient theory the static profile of the liquid density in the
  tube is  analytically obtained and compared with the profile issued from molecular dynamics simulation.  Inside the tube there are
  two domains: a thin layer near the solid wall where the liquid
  density is non-uniform and a central core where the liquid density
  is uniform.  In the dynamic case a closed form analytic solution
  seems to be no more possible, but by a scaling argument
 it is shown that, in the tube,  two distinct domains
  connected at their frontiers still exist. The thin inhomogeneous layer near the
  solid wall can be interpreted in relation with the Navier length
  when the liquid slips on the boundary as it is expected by
  experiments and molecular dynamics calculations.
\end{abstract}


\maketitle

\section{Introduction}
Nanofluidics is the study of the behavior of fluids that are confined
to structures of nanometer characteristic dimensions (typically 1-100
nm).  The possibility to observe liquids flowing at nano and micro
scales, for example in carbon nanotubes \cite{Iijima,Harris,Tabeling},
by using sophisticated experiments and complex molecular simulations
using Lennard-Jones forces reveals new behaviors that are often
surprising and essentially different from those usually observed at
macroscopic scale \cite{Ball,Rafii,Bonthuis}. For example, Majumder
\emph{et al} \cite{Majumder} perform some interesting experiments and
they estimate that, in nanotubes, the flow rates are four to five orders of magnitude
faster than conventional fluid flow predicted through pores of 7
nm diameter and, contrary to predictions based on classical
hydrodynamics, the flow rate does not decrease with increasing
viscosity.  Sinha \emph{et al}\cite{Mattia1}, in another set of
experiments, indicate that in carbon nanotubes ranging from  2 to 7
nm of diameter,  fluids flow with velocities up to 105 times faster than what
predicted by classical fluid dynamics calculations.

The critical dimension below which confinement in nanotubes  affects
fluid transport is currently debated.  For example if we consider
water molecules between two flat, hydrophobic surfaces, it has been
calculated\cite{Mattia2} that, at room temperature and atmospheric
pressure, this critical dimension is around $100$ nm.  Conversely,
some experiments seems to show that the continuum approximation
breaks down below $10$ nm in case of water, whereas experiments on
capillary filling of molten metals in $0.6 -1.2$ nm channels for
zeolites show that the threshold for confinement effects is closer to
$1$ nm\cite{Mattia2}.

These incongruences may be explained by the fact that actually there
is a severe computation limitation to molecular simulation, that the
smooth liquid-gas interface disappears in tubes with   diameter less
than $8-10$ nm and therefore anomalous behavior of water may be
observed in experiments with carbon nanotubes. Indeed, at this
nano-size, the surface chemistry and structure of nanotubes must be
controlled with a high precision to control flow rate and interaction
of fluid components\cite{Mattia2,Thomas}. Moreover, in the framework
of molecular dynamics, there are some problems to apply in a simple
and direct way the propest boundary conditions necessary to generate
the fluid flow. This is true especially when we consider pressure
driven flow\cite{Nicholls}.  Various methods exist to investigate fluid
transport in molecular dynamics. Examples are the \emph{gravitational
  field method}, where an \emph{artificial} gravitational force -- much
greater than the earth's gravitational pull -- is introduced or the
\emph{channel moving model}, a method to trigger the flow with the
viscous shear forces applied to the fluid by two moving channel walls.

Despite this indeterminacy in the literature, a relevant number of
experimental studies lead to the conclusion that the classical
Navier-Stokes equations are still valid at the nanoscale (see Bocquet
and Charlaix\cite{Bocquet} and included references).  The critical
threshold for the applicability of continuum hydrodynamics
investigated with molecular simulations and experiments may set around
$1$ nm.  \newline This value can be numerically obtained because
beyond the validity of continuum equations, the value of the viscosity quantitatively
remains  equal to the bulk value.  A typical correlation
time for the stress-stress correlation function is the picosecond
$\tau_{\sigma}=10^{-12}$ s, and the kinematic viscosity is $\nu =
10^{-6} \text{m}^2 \text{s}^{-1}$; consequently we obtain for water a
viscous length scale $\ell_c = \sqrt{\nu \tau_{\sigma} } \approx 1$
nm.  This observation seems to indicate, at least for water,
that an unexpected nano-metric characteristic length scale
naturally  emerges as the lowest bound for the validity of the notion of
viscosity.

The important conclusion, in analyzing the actual literature, is that for water under normal physicochemical conditions, the Navier-Stokes equation remains valid in nano-channels down to typically $1-2$ nm and   the discrepancy between molecular dynamics simulations and experiments seems to be induced  by the interaction of the fluid with the wall, \,i.e. when we consider the boundary conditions. The evidence of this conclusion is given by the measurements and the molecular dynamics simulations of the density profile which clearly fluctuates in the vicinity of a solid wall. Therefore the main problem is not if the continuum hypothesis has to be abandoned, but whence  the correct boundary conditions comes out.

Since van der Waals at the end of the $19$-th century, the fluid inhomogeneities in liquid-vapor interfaces are represented using
continuous models that allows to  take  account of a volume energy depending on space density derivative\cite{Dunn,Seppecher,widom,Kazm,Onuki}. Nevertheless, the corresponding square-gradient functional is unable to model repulsive force contributions and misses the dominant
damped oscillatory packing structure of liquid interlayers near a substrate wall\cite{Chernov}. Furthermore, the decay lengths
are correct only close to the liquid-vapor critical point where the damped oscillatory structure is subdominant\cite{Evans}. In mean field theory, weighted density-functional has been used to explicitly demonstrate the dominance of this structural contribution
in van der Waals thin films and to take  account of long-wavelength capillary-wave fluctuations as in papers that renormalize
the square-gradient functional to include capillary wave fluctuations\cite{Fischer}. In contrast, fluctuations strongly damp out
oscillatory structure and it is mainly for this reason that van der Waals  original prediction of a hyperbolic tangent profile is so close
to simulations and experiments\cite{Ono,Rowlinson}. It is possible to adjust, in phenomenological way, this state of affairs by considering the approach by Cahn in his celebrated paper studying wetting near a critical point \cite{Cahn0}. An  approach that may be justified via  a suitable asymptotic expression considering approximation of hard sphere molecules and London potentials for liquid-liquid and solid-liquid interactions\cite{Gouin 2}: in this way, we took account of the power-law behavior which is dominant in a thin liquid film in contact with a solid.

It is found that  a similar situation may be also considered  for the flow of the fluids and not only for their densities. The amended boundary conditions at a solid surface in the nano-scale framework must introduce a new length, the so-called \emph{Navier length} or \emph{slip length}\cite{Blake,Landau,Navier}: a length relating the tangential velocity  to the shear rate at the wall. Liquid slip is essential in nano-fluidic systems, as shrinking channel size leads to a dramatic increase of flow resistance and thus high-energy consumption for driving nonslip flow\cite{MH,Ma}.

The aim of the note is to justify the boundary conditions equations of nano-fluid mechanics using a simple mesoscopic approach. Our basic idea  has been suggested by some experimental work regarding the measurement of the density of water in narrow pores \cite{Ball,Bear}. In such experiments  it is shown that at the nanoscale the liquid must be compressible and inhomogeneous  in a very narrow layer near the solid wall. In our opinion this layer is connected with the Navier length.

To support this idea we consider a nanotube made up of a cylindrical hollow tube whose diameter is of some nanometers.  The nanotube is immersed in a liquid filling the interior of the nanotube, and to take   account of the compressibility of the liquid, we use a second
gradient theory in which the fluid is modeled by a van der Waals fluid for which the surdeformations are taken into account\cite{van der Waals,Korteweg,Cahn,Isola}.  Therefore,  we use a continuum theory in which
the volume  energy of the liquid is a function not only of the
density but also of the gradient of density. The associated mathematical model
may be obtained via a molecular mean field theory \cite{Gouin 1,Gouin 3} or via the axiomatic theory of the thermomechanics of continua \cite{Forest} or by
considering maximization of the entropy production\cite{Raja1,Raja2}.  In the following, the ideas of the van der Waals square gradient functional is used together with a condition at the wall taking account of the fluid density at its immediate proximity \cite{Gouin 2,Gouin 3}. By using this \emph{continuum} approach, we provide a bridge between classical models of fluid mechanics and molecular simulations. A  framework to develop simple analytical results in closed form of technical importance.

The plan of the paper is the following. Section 2 is devoted to the
basic equations of capillary fluids using a second gradient theory. In
Section 3 we consider the static  problem to obtain the density
profile of the liquid in the nanotube and its comparison with molecular dynamics simulation. In Section 4 we consider a
dimensional analysis of Navier-Stokes equations in cylindrical
coordinates and we show that the inhomogeneous character of the
governing equations introduce in a \emph{natural} way the Navier
length.  The last section is devoted to  remarks and conclusion.
\vfil\eject

\section{Capillary fluids}
\subsection{Basic equations}

Let us consider a fluid in a nanotube. In the immediate vicinity of
the solid wall of the nanotube, the intermolecular forces are dominant
and the density profile of the confined fluid is
inhomogeneous;  in the case of a small variation of density,
the intermolecular forces
induce  a sharp variation of the gradient of   density at the wall. In this framework   the specific fluid internal
energy $\varepsilon$, which is usually a function only of the density $\rho $ and the specific entropy $s$,
must also  take  account of the gradient of   density $\func{grad}\rho$.

The \emph{second gradient} model \cite{Germain} is a
 theory of continua based on  constitutive
equations depending on the gradient of the density.  In this case,
restricting first our attention to statics, we start from a specific internal
energy density in the form
\begin{equation*}
\varepsilon =f(\rho, s, \beta ) \quad \mathrm{{with}\quad }\beta =(\func{grad%
}\rho )^{2},
\end{equation*}

\noindent
 and in such a way the  stress tensor is \cite{Gouin 1}
\begin{equation}
\label{3a}
\mathbf{\mathbf{\sigma }}=  -p\,\bm{I}-\lambda \,(\func{grad}\rho )\otimes(\func{%
grad}\rho )  \equiv -p\,\bm{I}-\lambda \,(\func{grad}\rho )(\func{%
grad}\rho )^{T}\ %\ \mathrm{{or}\ }\ \sigma _{ij}=-p\ \delta _{ij}-\lambda
%\,\rho _{,i}\rho _{,j}\,,\ i,j\in \{1,2,3\}
 \end{equation}%
 where $\lambda\equiv 2\,\rho \,\varepsilon _{\beta }^{\prime }\ ,%
 \ p\equiv \rho ^{2}\varepsilon _{\rho }^{\prime
 }-\rho\,\func{div}(\lambda \,\func{%
   grad}\rho )$ is the spherical part of the stress tensor, $\bm{I}$ is the identity tensor and $\,^T$ denotes the transposition.  \newline The
 scalar $\lambda $ - call the surdeformation coefficient of the fluid - accounts for surdeformation effects and generally
 depends on $\rho, s$ and $\beta $.  By using kinetic theory, Rowlinson and Widom
 \cite{Rowlinson} obtained an analogous result
  but with $\lambda$ constant at a given temperature $T$ and the
 specific energy $\varepsilon $ reads
\begin{equation*}
\rho\, \varepsilon (\rho ,s,\beta )=\rho\, \alpha (\rho ,s)+ \frac{\lambda}{2}  \, \beta ,
\end{equation*}%
where $\alpha
(\rho ,s)$ is the the specific internal energy of the classical compressible fluid of pressure   $P \equiv \rho ^{2}\alpha _{\rho }^{\prime }$ and
temperature   $T \equiv\alpha _{s}^{\prime }$. Consequently, in Eq. (\ref{3a}),
\begin{equation*}
p=P-\lambda \left(\frac{\beta }{2}+\rho\, \Delta \rho\right) \quad {\rm and},
\end{equation*}
\begin{equation*}
\mathbf{\mathbf{\sigma }} = -P\,\bm{I}+\lambda \,\left(\, \frac{1}{2}\ \left((\func{grad}\rho )^2 +\rho \, \Delta \rho \right)\, \bm{I} - (\func{grad}\rho )(\func{%
grad}\rho )^{T}\right),
\end{equation*}
where $\Delta$ denotes the Laplacian operator.
Because a convex equation of state is not able to connect the different bulks associated with a fluid interface,   many  authors use the van der Waals equation of state or other similar laws for the thermodynamical pressure $P$.
In fact, we  only consider the liquid bulk and the thermodynamical pressure $P$ is expanded near the bulk density. The equation of
motion is
\begin{equation}
\rho \,\mathbf{a} = \func{div} \mathbf{\mathbf{\sigma }}- \rho \func{grad}\Omega ,  \label{motion1}
\end{equation}
where  $\Omega$ is the extraneous force potential. Let us denote $\omega =\Omega -\lambda \,\Delta \rho $, then the
equation of motion yields \cite{Gouin 1}
\begin{equation*}
\rho \,\mathbf{a}+\func{grad}P+\rho \func{grad}\omega =0.  \label{motion2}
\end{equation*}%
This relation is similar to the one of the perfect fluid case but the term
$\omega $ involves all capillarity effects. %
By neglecting the extraneous force potential, we obtain
\begin{equation}
\rho\,\mathbf{\mathbf{a}}+\func{grad}P-\lambda \,\rho \,\func{grad}%
\Delta \rho = 0\label{motion3}.
\end{equation}
The equation of motion (%
\ref{motion3}) can also be written in the form \cite{Gouin 4}
\begin{equation*}
\mathbf{a} =T \func{grad}s-\func{grad} H  ,  \label{thermotion}
\end{equation*}
and if $T$ is constant,
\begin{equation}
\mathbf{a} +\func{grad} \pi  =0 \,,  \label{thermotion1}
\end{equation}
with the potentials
\begin{equation*}
  H =  \varepsilon+ \frac{p}{\rho} \equiv h - \lambda\, \Delta  \rho
  \quad {\rm and}\quad   \pi  =
  H-T\,s \equiv \mu- \lambda\, \Delta   \rho
\end{equation*}
being the \emph{generalized enthalpy and generalized chemical potential}
of the capillary fluid, respectively, where
\begin{equation*}
  h = \alpha +  \frac{P} {\rho}  \quad {\rm and}\quad   \mu = \alpha +
  \frac{P}{\rho} - T\,s
\end{equation*}
are the enthalpy and the chemical potential of the classical
compressible fluid, respectively \cite{Gouin 4}.

In the case of viscous fluids, the equation of motion takes
account of the viscous stress tensor  which is classically given by
\begin{equation*}
  \bm{\mathbf{\sigma }}_{v}=\eta ( \mathtt{tr}\,\bm{D}) \bm{I}+2\,\kappa \,\bm{D},
\end{equation*}
 where $\eta$ and $\kappa$ are the shear and bulk viscosity
coefficients respectively assumed to be constant and
%for our purpose it is possible to use
%the Stoke's hypothesis $3\, \eta+ 2\, \kappa = 0$ \cite{Sli};
$\bm D$
is the deformation tensor, symmetric gradient of the velocity field \cite{Sli}. It would be coherent to add terms
accounting for the influence of higher order derivatives of the
velocity field but the over-deformation only comes from the
density. In fact,  as discussed in
introduction,
the Navier-Stokes equations correctly take   account of the
 viscous behavior  without  higher order derivatives of the
velocity field.
Equation (\ref{motion1}) is modified as $ \rho
\,\mathbf{a}=\func{div}(\mathbf{\mathbf{\sigma }}+\mathbf{\mathbf{%
    \sigma }}_{v})$ and for viscous liquids, Eq. (\ref{motion3}) writes
\begin{equation}
  \rho \,\mathbf{a}+\func{grad}P-\lambda\,\rho \func{grad} \,\Delta
  \rho -\func{div}\mathbf{\mathbf{\sigma }}_{v}=0.  \label{viscous_motions}
\end{equation}

\subsection{Boundary conditions}

The forces acting between liquid and solid range over a few nanometers
but can be simply described by a special surface energy.  For a solid
wall, the total surface   energy $\varphi$ at the wall is expressed as \cite{De
Gennes2}:
\begin{equation}
  \varphi (\rho_{_S})=-\gamma _{1}\rho_{_S} +\frac{1}{2}\,\gamma
  _{2}\,\rho_{_S}^{2}.  \label{surface energy}
\end{equation}%
Here $\rho_{_S}$ denotes the limit value of the liquid density at the
solid wall; the constants $\gamma _{1}$, $\gamma _{2}$ and
$\lambda $ are positive and can be obtained by the mean field
approximation in molecular theory \cite{Gouin 2}.    The boundary
condition for the liquid density at the solid wall $(S)$ is associated
with the free surface energy (\ref{surface energy}) and was calculated
in \cite{Gouin 3}
\begin{equation}
  \lambda \left( \frac{d\rho }{dn}\right) _{|_{S}}+\varphi ^{\prime
  }(\rho_{_S})\ =0,  \label{cl1}
\end{equation}%
where $\displaystyle \frac{d}{dn}\ $  means the derivative following the direction of the external normal $\textbf{n}\,$ to the fluid. This condition corresponds to an embedding effect for the density of the fluid which is not taken into account in classical hydrodynamics.

The aim of the present note is to show that the boundary condition
\eqref{cl1} introduce a nano-boundary layer in the tube. The byproduct
of this layer is the presence of a \emph{slip} velocity that we read
at the micro scale and this also when we consider the classical
no-slip boundary condition for the velocity field at the wall.

\subsection{The chemical potential in the liquid phase}

Due to the fact $ \mu$ is defined to an additive constant, we denote
by $\mu _{_{0}}(\rho )$ the chemical potential of the fluid for the liquid-vapor plane interface, such that
\begin{equation*}
\mu _{_0}(\rho_l)=0,
\end{equation*}
where $\rho _{l}$ is the liquid density in the liquid bulk
corresponding to the plane liquid-vapor interface at a given
temperature $T$.  \newline To the liquid bulk of density $\rho
_{l_b}\neq \rho _{l}$ - the density $\rho _{l_b}$ does not correspond
to a plane liquid-vapor interface but to a \emph{mother} liquid bulk
associated with a droplet or a bubble and does not verify the Maxwell
rule of equal area corresponding to plane liquid-vapor interfaces
\cite{Derjaguin} - we associate $\mu _{l_{b}}(\rho )\equiv\mu
_{_{0}}(\rho )-\mu _{_{0}}(\rho _{l_b})$ corresponding to the chemical potential for the mother liquid bulk $\rho _{l_b}$.  The thermodynamical
potentials $\mu _{l_b}$ can be expended at the first order near the
liquid bulk of density $\rho _{l_b}$
\begin{equation*}
\mu _{l_{b}}(\rho )=  \frac{c_{l}^{2}}{\rho _{l}} \left( \rho -\rho _{l_b}\right),
\end{equation*}
where $c_l$ is the isothermal sound velocity in the liquid bulk of
density $\rho_l$ \cite{Gouin 7}. Similarly, the thermodynamical
pressure is expended as
\begin{equation}\label{presure}
P = P_{l}+ c_l^2 \left( \rho -\rho _{l}\right),
\end{equation}
where $P_{l}$ is the thermodynamical pressure in the liquid bulk of
density $\rho_l$.

\section{Liquid density in a nanotube at equilibrium}

A nanotube is represented by a hollow cylinder of length size $L$ and
of small diameter $d=2R$, ($d/L\ll 1$). In Subsection IIIA, $d$ ranges from 2 to 100 nanometers and $L$ is of the order of some microns.
\subsection{Profile of density by using the continuum approach}
We consider solid walls with a
large thickness with regards to molecular dimensions such that the
surface energy verifies an expression in form (\ref{surface energy}).
At equilibrium ($\mathbf{a}=0$), far from the nanotube tips and by neglecting the
external forces ($\pi =\mu
_{_{0}}-\lambda \,\Delta \rho $), Eq. (\ref{thermotion1})  implies the profile of density as solution of the differential equation
      :
\begin{equation*}
  \lambda \,\Delta \rho =\mu _{_{0}}(\rho )-c,  \label{densityequ}
\end{equation*}%
where $c=\mu _{_{0}}(\rho _{l_b}) $ is an additional constant associated with the density value
$\rho _{l_b}$  in the mother bulk  outside the nanotube    \cite{Derjaguin}.  We consider the case when only the
liquid fills up the nanotube. The profile of density is given by the
differential equation :
\begin{equation}
  \lambda \,\left(  u_{rr} +\frac{1}{r}\, u_r\right) -%
  \frac{c_l^{2}}{\rho _l}\ u=0,\qquad \mathrm{with}\quad u=\rho
  -\rho _{l{_b}}.  \label{densityprofile1}
\end{equation}%
In cylindrical coordinates, $r$ denotes the radial coordinate. The
reference length is
\begin{equation*}
\delta _{l}=\sqrt{\frac{\lambda \,\rho{_l}}{{c_l}^{2}}}\,.
\end{equation*}%
We denote by $x$ the  dimensionless variable such that $r=\delta
_l\,x$. Equation (\ref{densityprofile1}) reads
\begin{equation}
u_{xx}+\frac{1}{x}\,u_x-\, u=0.
\label{densityprofile2}
\end{equation}%
The solutions of Eq. (\ref{densityprofile2}) in classical
expansion form $u=\sum_{n=0}^{\infty }a_{n}x^{n}$  yield
\begin{equation*}
\sum_{n=2}^{\infty }n^{2}\,a_{n}\,x^{n-2}-a_{n-2}\,x^{n-2}=0\quad
\Longrightarrow \quad n^{2}\,a_{n}=a_{n-2}\,.
\end{equation*}%
Due to the symmetry at $x=0$, the odd terms are null and consequently,
\begin{equation*}
u=a_{_{0}}\,\sum_{p=0}^{\infty }\ \frac{1}{4^{p}\,(p\,!)^{2}}\ x^{2p}\,.
\end{equation*}%
The series has an infinite radius of convergence. Let us define the
functions
\begin{equation*}
  f(x) \equiv \sum_{p=0}^{\infty }\ \frac{1}{4^{p}\,(p\,!)^{2}}\ x^{2p}
 \qquad
\emph{and} \qquad  g(x)  \equiv f^{\prime }(x)=\sum_{p=1}^{\infty }\ \frac{2p}{%
4^{p}\,(p\,!)^{2}}\ x^{2p-1}.
\end{equation*}%
Consequently, $u=a_{_{0}}\,f(r/\delta_l)$. The boundary condition (%
\ref{cl1}) at $x=R/\delta_l$ yields
\begin{equation*}
\frac{\lambda}{\delta_l} \,\frac{du}{dx}=  \gamma _{1}-\gamma _{2}\,\rho
  \qquad {\rm or} \qquad a_{_{0}}=\frac{\delta
_l\,\left( \gamma _{1}-\gamma _{2}\,\rho _{l_b}\right) }{\lambda
\,g\left( \frac{R}{\delta _l}\right) +\gamma _{2}\,\delta
_l\,f\left( \frac{R}{\delta _l}\right) }
\end{equation*}%
and the density profile reads
\begin{equation*}
\rho =\rho _{l_b}+\frac{\delta_l\,\left( \gamma _{1}-\gamma
_{2}\,\rho _{l_b}\right) }{\lambda \,g\left( \frac{R}{\delta_l}%
\right) +\gamma _{2}\,\delta _l\,f\left( \frac{R}{\delta _l}%
\right) }\ f\left( \frac{r}{\delta _l}\right) .\label{densityprofile}
\end{equation*}
% In fact, practically, $\rho _{l_b}\simeq \rho _{l}$ and the profile of
% density can be written as
Densities $\rho _{l_b}$ and $\rho _{l}$ differ very slightly and, for
the purposes of this work, can be considered as coinciding. Finally, the density profile can be written as
\begin{equation}
  \label{density_profile}
  \frac{\rho}{\rho_{l}}=1+\frac{\gamma_1-\gamma_2 \rho_l}{\delta_l c_l^2
    g\left(\frac{R}{\delta_l}\right) +\gamma_2 \rho_lf\left(\frac{R}{\delta_l}\right)}f\left(\frac{r}{\delta_l}\right)
\end{equation}

In order to visualize the density profiles (\ref{density_profile}) we
consider the case of the water at $20^\texttt{o}$ Celsius, for which the different physical
constants involved in the model are (in \textbf{cgs} units) as follow :\newline
$\rho_l=0.998$, $c_l=1.478 \times 10^5$ and $\lambda=1.17 \times
10^{-5}$; the value of $\gamma_2$  only depends on the fluid and in the
case of  water  $\gamma_2=54$ , whereas the
coefficient $\gamma_1$ is related to the hydrophobicity or to the
hydrophilicity of the solid wall\,\cite{Gouin 7}.

\begin{figure}
  \begin{center}
    \includegraphics[width=6.4cm]{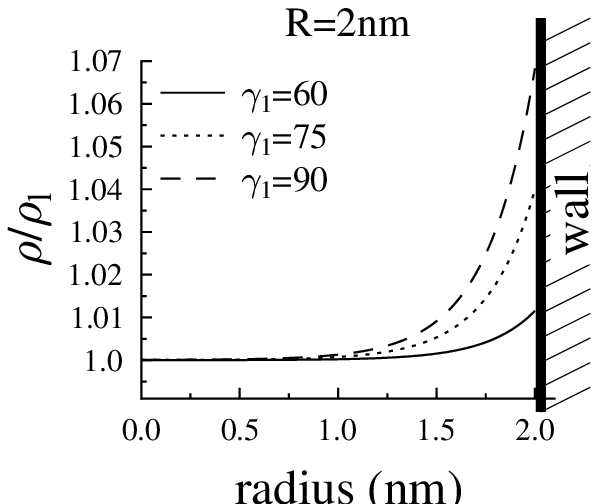}\qquad
    \includegraphics[width=6.4cm]{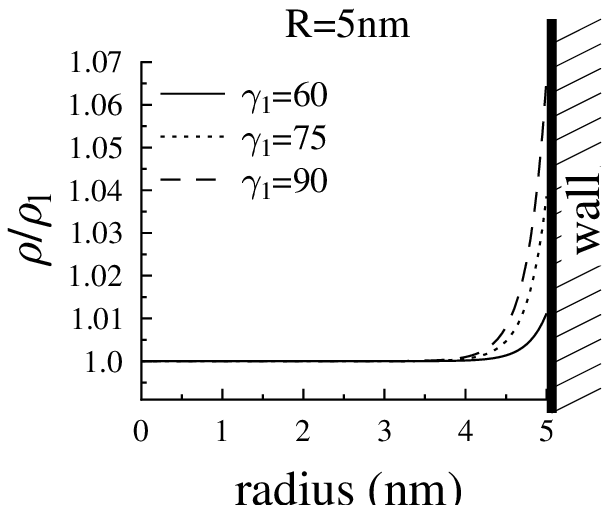}\\[0.1cm]
a)\hskip 7.2cm b)\\[0.3cm]
    \includegraphics[width=6.4cm]{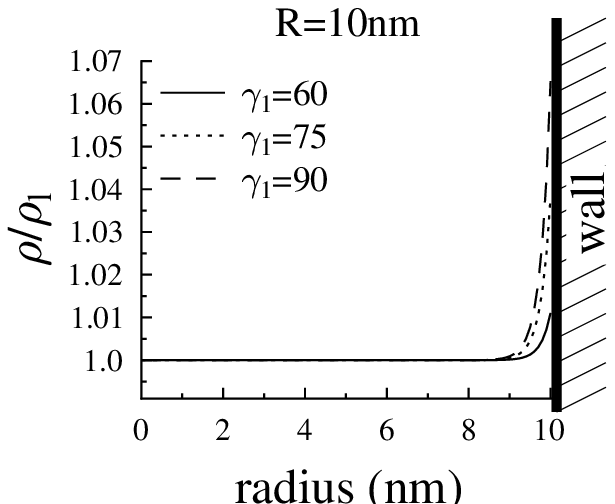}\qquad
    \includegraphics[width=6.4cm]{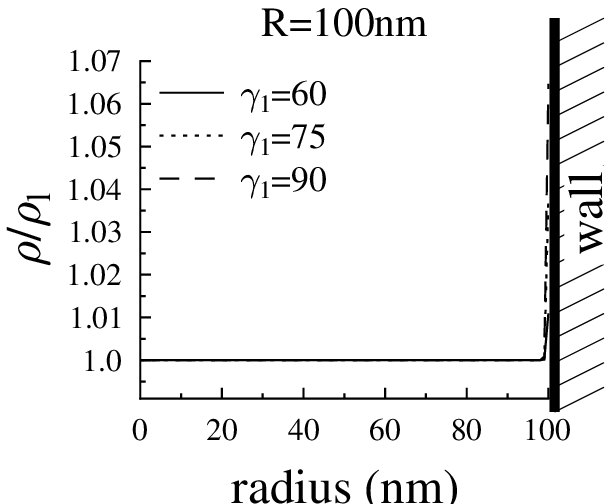}\\[0.1cm]
c)\hskip 7.2cm d)
  \end{center}
  \caption{Density profiles for different values of $\gamma_1$
    ($\gamma_1=60$, $75$, $90$) inside four nanotubes of different radius:
    (a) $R=2$ nm, (b) $R=5$ nm, (c) $R=10$ nm, (d) $R=100$ nm. Whatever
    the radius of the tube is, we note that the main part of the
    density variation near the wall accounts for a thickness of one
    nanometer and  that, outside   this thin layer, the liquid
    density is strongly constant.}\label{fig1}
\end{figure}

In Figure \ref{fig1}, different density profiles obtained
for four tubes of radius $R=2,5,10$ and $100$ nanometers and for different values of $\gamma_1$ ($\gamma_1=60,
75, 90$), corresponding to the case when the solid wall is
hydrophilic  are plotted.
The density profiles plotted in Figure \ref{fig1} show that at
equilibrium and independently of the diameter of the tube, the fluid
domain can be separated in two cylindrical domains: the \emph{core} in
the center of the tube, where the density is constant, and the
\emph{boundary layer} near the solid wall of the tube, where  the
gradient of the density is significant.  The thickness of the layer
wherein the variation of the density takes place, is about four times
the value of $\delta_l=0.231$\,nm.

The maximal value of the density is reached on the boundary (at
wall-fluid interface). It depends both on the value of the coefficient
$\gamma_1$ and, to a lesser extent, on the diameter of the tube. The
density variation inside the tube is moderate: at most $6.8\, \%$
for a strongly hydrophilic wall ($\gamma_1=90$) and for a tube of
tiny radius $R=2$ nm.
\subsection{Comparison between continuum approach and molecular dynamics simulation}

\begin{figure}
  \begin{center}
    \includegraphics{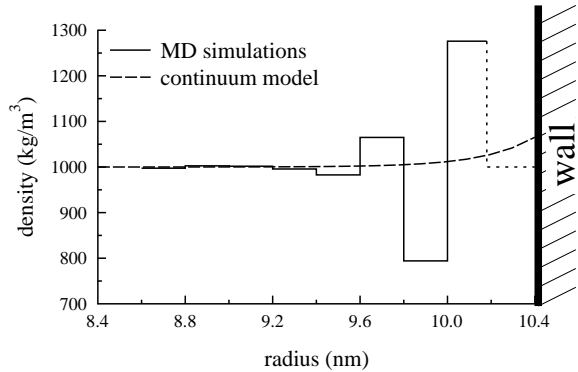}
  \end{center}
  \caption{Density profiles inside a nanotube of radius 10.4 nm issued
    from molecular dynamics simulation (continuous line) and from the
    continuum model (dashed line). The dotted line which extends the
    density profile issued from MD simulation is associated with the
    gap of density du to the lack of water molecules beside the wall.}
    \label{fig2}
\end{figure}

Molecular dynamics (MD) simulations take account of van der Waals
forces by using Lennard-Jones interaction potentials between a small
number of molecules included inside the nanotube. Near the wall, MD
simulations show oscillatory density profiles corresponding to the
variations of the indicator function of molecular presence; moreover,
the non-penetrability condition of the water molecules leads to empty
domains beside the wall\cite{Mattia2,Sony}.  These density
fluctuations are obviously in contrast with the predictions of
continuum studies corresponding to an averaging in molecular
energies. In the layer beside the wall of about one nanometer, MD
simulations consider a few number of molecules.  As pointed out by
Thomas and McGaughey \cite{Thomas2} (in Fig. 3 and Fig. 4), the graphs of
density near the wall are not associated with continuous functions;
the molecular distributions are gathered in cylindrical layers of
about $0.2$ nm of thickness and the continuous guidelines are simply
added between the density values of cylindrical layers to highlight
the minima and maxima of the layer densities. Consequently, the
comparison between MD simulations and the continuum approach
corresponding to an averaging of the sum of molecular potentials must
be done on the Gibbs adsorption of density \cite{Rowlinson} at the
nanotube wall involving the domain where the density differs from the
bulk density. Our comparison is done by reference to the examples
presented in the paper by Thomas and McGaughey. The density profile
retained for comparison purpose is plotted in Figure \ref{fig2}.

In continuum theory of capillarity, the Young angle $\theta$ between
solid-liquid surface and liquid-vapor interface is given by the
relation:
\begin{equation}
\sigma_{_{SV}}-\sigma_{_{SL}} =  \sigma_{_{LV}}  \cos \theta , \label{young}
\end{equation}
where $\sigma_{_{SV}}, \sigma_{_{SL}}, \sigma_{_{LV}}$ are
respectively the solid-vapor, solid-liquid and liquid-vapor
superficial tensions.  For water at $20^{\circ}$ Celsius and in
\textbf{cgs} units, $\sigma_{_{LV}} \simeq 72$ and $\sigma_{_{SV}}$
can be neglected. Relation (\ref{surface energy}) expresses the value of
$\sigma_{_{SL}}$ by mean-field theory in capillarity ($\sigma_{_{SL}}=\varphi(\rho_{_S})$).\newline
Using a mean field model and London
forces \cite{Gouin 2} the $\gamma_2$ value for water is obtained in\ \cite{Gouin 7} and reads $\gamma_2 \simeq
54$. Consequently, from Eqs. (\ref{surface energy}) and (\ref{young}),
$\gamma_1 \simeq 96$ and $\gamma_1 \simeq 75$ correspond to a Young
angle of $0$ degree and $45$ degree, respectively. Mattia and Gogotsi
give a range of values of the Young angle for graphite
\cite{Mattia2}. A realistic value for carbon nanotube can be taken as
$\gamma_1 \simeq 90$.
\newline
As a relevant example for nanotubes, the
graphs of density associated with the MD simulations and continuum
model are presented on Fig.  \ref{fig2}.  The MD simulation profile is
rebuilt from Fig. 3 in \cite{Thomas2}, where guidelines added between
minima and maxima of densities are replaced by a step function
corresponding to the cylindrical layers shown in Fig. 4 in
\cite{Thomas2}.  Both profiles of density, corresponding to
the two models, differ from the uniform bulk density value only in the
nanometer range near the wall. In this domain, we calculate the total
mass for the MD simulation  as well as  for the continuum model;
consequently, we are able, in the two cases, to compare the Gibbs
adsorption at the wall. To take account of the gap of density near the
wall appearing in MD simulations, the cylindrical layer near the wall
associated with MD simulation is considered in size 10 per cent
smaller than the other layers. For carbon nanotube with radius of
10.4 nm, MD simulation predicts a Gibbs adsorption per unit length at
the wall of $11.4\times 10^{-15}$ g cm$^{-1}$ whereas the continuum
model predicts a Gibbs adsorption per unit length at the wall of
$9.7\times 10^{-15}$ g cm$^{-1}$. These two values are of the same
order. Considering that the water molecule mass is about $3\times
10^{-23}$ g, we obtain a Gibbs adsorption of about 30 molecules per
nanometer length of the nanotube.

In Table \ref{tab_MD} are shown the values of the Gibbs adsorption at
the wall predicted by the continuum model for different values of the
parameter $\gamma_1$. We observe that complete similarity between the
two models is obtained for the perfect wetting.

\begin{table}[b]
  \centering
  \begin{tabular}{|c||c|c|c|}
\hline\hline
$\gamma_1$ & 75 & 90& 96 \\
\hline

Gibbs adsorption (g cm$^{-1}$)& $5.69\times 10^{-15}$ & $9.74\times 10^{-15}$ & $11.36\times
10^{-15}$  \\
\hline\hline
\end{tabular}
\vskip 0.15cm

\caption{ Gibbs adsorption at the wall predicted by the continuum model for different
values of the parameter $\gamma_1$}
\label{tab_MD}
\end{table}

\noindent We can conclude:\\
In the two models we obtain the same thickness of the domain where
the density of water is different from the bulk
density.\\
The Gibbs adsorption at the wall is similar for the
two models.\\
In the comparison, the continuous mean-field
theory uses London potential which is an approximation of
Lennard-Jones potential but the difference of Gibbs adsorption between
the two models is, in this example, less important than the disparity
between the MD simulation results obtained in different papers in the
literature \cite{Majumder,Nicholls,Rafii,Sony}.

\section{Motion of liquid in a nanotube}

Due to the cylindrical symmetry of the problem, it is supposed that
the velocity field $\vec{v}$ and the density $\rho$ have a radial
symmetry
\begin{equation*}
\vec{v}=u(r,z) \vec{e}_r+w(r,z)\vec{e}_z,\qquad\rho=\rho(r,z),
\end{equation*}
where $(\vec{e}_r, \vec{e}_\theta, \vec{e}_z)$ is the basis of the
cylindrical coordinates $(r,\theta, z)$. The continuity equation is
then written as
\begin{equation}
\frac{1}{r}(r \rho u)_r+(\rho w)_z=0 . \label{masscyl}
\end{equation}
 In the following,
and only for the sake of algebraic simplicity, %we assume
Stokes' hypothesis concerning the viscosity
is assumed : $3\, \eta+ 2\,\kappa=0$.
%For all the results, we find that
This assumption is not essential,
%with an important difference with the general case,
but the analytic
development is simplified and the comprehension of calculations is
easier.  %more straightforward.
\newline
In the steady case $( {\partial
  \vec{v}}/{\partial t}=0)$, the non-vanishing equations of motion
(\ref{viscous_motions}) are written as
\begin{eqnarray}
\rho \left(u u_r+w u_z \right)&=&-P_r+\kappa
  \left\{\frac{4}{3}\,\left[\frac{1}{r}\left(ru\right)_r\right]_r
    +u_{zz}+\frac{1}{3}\,w_{rz} \right\}\nonumber\\
& &+ \lambda \rho \left[ \frac{1}{r} \left(r \rho_r
    \right)_r+\rho_{zz} \right]_r,\label{momentum1} \\
\rho \left(u w_r+w w_z\right)&=&-P_z+\kappa
  \left\{\frac{1}{r}\left(rw_r\right)_r+\frac{4}{3}\,w_{zz}+\frac{1}{3}\left[
      \frac{1}{r} \left(r u \right)_r\right]_{z} \right\} \nonumber\\
& &+ \lambda \rho \left[
    \frac{1}{r} \left(r \rho_r \right)_r+\rho_{zz} \right]_z. \label{momentum2}
\end{eqnarray}
The solution of this set of equations cannot be obtained
analytically. However, an approached velocity profile can be obtained
by re-scaling   Eqs (\ref{masscyl}--\ref{momentum2}). The re-scaling
procedure, which is the object of the present section, is made with
respect to a small geometrical parameter $\epsilon=d/L$   but also with
respect to a small physical quantity $\tau=\delta_l/d$. To this goal,
the following set of dimensionless  variables -- indicated with\ \ $\tilde{ }$\ \ --\ \
is introduced :
\begin{equation*}
r=d \tilde{r}, \quad z= L \tilde{z}, \quad u= \hat{w} \tilde{u}, \quad
w= \hat{w} \tilde{w}, \quad \rho=\rho_l\tilde{\rho},
%\quad           {\rm{with}} \quad\hat{\rho}\equiv\rho_l ,
\end{equation*}
where $\hat{w}$ is a reference velocity of the liquid; we chose
the mean velocity in the nanotube
estimated by its corresponding value in the case of a  Poiseuille flow
\begin{equation}
\hat{w} = - \frac{d^2\,\func{grad} \Delta P}{32\, \kappa},\label{meanw}
\end{equation}
where $\func{grad}\Delta P$ denotes the gradient of the pressure
difference between the nanotube extremities.
In so doing, the continuity equation becomes
\begin{equation}
(\tilde{r} \tilde{\rho} \tilde{u})_{\tilde{r}}+\epsilon\, \tilde{r} (\tilde{\rho} \tilde{w})_{\tilde{z}}=0.\label{massd}
\end{equation}
If we denote by $Re = \rho_l\,\hat{w}\, d/\kappa$ the Reynolds number and by $M = \hat{w}/c_l$ the Mach number,  and taking account of Eq.(\ref{presure}),
 the momentum equations become
\begin{eqnarray}
Re\, \tilde{\rho} \left(\tilde{u} \tilde{u}_{\tilde{r}}+ \epsilon \tilde{w} \tilde{u}_{\tilde{z}} \right)&=&-\frac{Re}{M^2} \, \tilde{\rho}_{\tilde{r}}+ \frac{4}{3}\left[\frac{1}{\tilde{r}}\left(\tilde{r}\tilde{u}\right)_{\tilde{r}}\right]_{\tilde{r}}+\epsilon^2 \tilde{u}_{\tilde{z}\tilde{z}}+\frac{1}{3}\,\epsilon\tilde{w}_{\tilde{r}\tilde{z}}\nonumber\\
& & + \frac{Re}{M^2}\, \tau^2 \tilde{\rho} \left[ \frac{1}{\tilde{r}} \left(\tilde{r} \tilde{\rho}_{\tilde{r}} \right)_{\tilde{r}}+\epsilon^2 \tilde{\rho}_{\tilde{z}\tilde{z}} \right]_{\tilde{r}},\label{moment1}
\\
Re\, \tilde{\rho} \left(\tilde{u} \tilde{w}_{\tilde{r}}+\epsilon \tilde{w} \tilde{w}_{\tilde{z}} \right)&=&-\frac{Re}{M^2}\, \epsilon \tilde{\rho}_{\tilde{z}}+\frac{1}{\tilde{r}}\left(\tilde{r} \tilde{w}_{\tilde{r}}\right)_{\tilde{r}}+\frac{4}{3}\, \epsilon^2 \tilde{w}_{\tilde{z}\tilde{z}}+\frac{1}{3}\, \epsilon \left[ \frac{1}{\tilde{r}} \left(\tilde{r} \tilde{u} \right)_{\tilde{r}}\right]_{\tilde{z}}\nonumber   \\
& & + \frac{Re}{M^2}\, \epsilon \tau^2 \tilde{\rho}\left[ \frac{1}{\tilde{r}} \left(\tilde{r} \tilde{\rho}_{\tilde{r}} \right)_{\tilde{r}}+\epsilon^2 \tilde{\rho}_{\tilde{z}\tilde{z}} \right]_{\tilde{z}}.\label{moment2}
\end{eqnarray}

In order to evaluate the respective size of the
coefficients of Eqs. (\ref{massd}--\ref{moment2}) some numerical reference
values for different physical variables should be considered. These
numerical values  are expressed in \textbf{cgs} units  as follows :
\begin{equation*}
L = 10^{-2},\quad c_l = 1.478 \times 10^5, \quad \kappa = 0.01,\quad \rho_l = 0.998,\quad \delta_l = 2.31 \times 10^{-8},
\end{equation*}
 and nanotubes of four different diameters are considered :
\begin{equation*}
d\in\left\{4\times 10^{-7},\quad 10^{-6},\quad 2 \times10^{-6},\quad 2 \times10^{-5}\right\}.
\end{equation*}
 We will assume
$\func{grad}\Delta P= -10^6$ (corresponding to one atmosphere per centimeter length of the nanotube).
Consequently, the numerical values of the coefficients in equations
(\ref{massd}--\ref{moment2}) are resumed in  Table \ref{TableKey1}.
\begin{table}[tbp]
\centering
\begin{tabular}{|c|c|c|c|c|c|c|c}
\hline\hline
d\,=\,2\,R & $4\times 10^{-7}$ & $10^{-6}$ & $2\times 10^{-6}$ & $2\times  10^{-5}$  \\
\hline\hline
$\epsilon$ & $4\times 10^{-5}$ & $10^{-4}$ & $2\times 10^{-4}$ & $2\times 10^{-3}$\\
\hline
$\tau$ & $5.8\times 10^{-2}$ & $2.3\times 10^{-2}$ & $1.16\times10^{-2}$ & $1.16\times 10^{-3}$\\
\hline
$\hat{w}$ & $5\times 10^{-7}$ & $3.13\times 10^{-6}$ & $1.25\times 10^{-5}$ & $1.25\times 10^{-3}$\\
\hline
%$Re$ & $5\times 10^{-7}$ & $3.13\times 10^{-6}$ & $1.25\times 10^{-5}$ & $1.25\times 10^{-3}$\\
$Re$ & $2\times 10^{-11}$ & $3.12\times 10^{-10}$ & $2.5\times 10^{-9}$ & $2.5\times 10^{-6}$\\
\hline
$M$ & $3.38\times 10^{-12}$ & $2.11\times 10^{-11}$ & $8.46\times 10^{-11} $ & $8.46\times 10^{-9}$\\
\hline
% $\xi$ & $0.5$ & $0.2$ & $0.1$ & $10^{-2}$\\  \hline
% ${Re}\,\xi /M^{2}$ & $2.18\times 10^{16}$ & $1.40\times 10^{15}$ & $1.74\times
% 10^{14}$ & $ 1.74\times 10^{11}$\\
 ${Re}\,\epsilon /M^{2}$ & \multicolumn{4}{|c|}{$6.98\times 10^{7}$}\\
\hline
$\xi=4\tau$ & $0.23$ & $0.092 $ & $0.046$ & $0.0046$\\  \hline
${Re}\,\epsilon\, \xi^2 /M^{2}$ & $3.73\times 10^{6}$ & $5.97\times
10^{5}$ & $1.49\times
10^{5}$ & $ 1.49\times 10^{3}$\\
\hline\hline
\end{tabular}
\vskip 0.15cm
\caption{Numerical values of the coefficients in equations
(\ref{massd}--\ref{moment2}) }
\label{TableKey1}
\end{table}

It is worth noting that the coefficient ${Re}\,\epsilon /M^{2}$ is
independent of the diameter of the nanotube. Moreover,  when $\func{grad}\Delta
P= -1$  corresponding to a very low pressure difference between
the tips of the nanotube, the term    ${Re}\,\epsilon /M^{2}$ is simply
multiplied by $10^{-6}$ which always remains  very large with
respect to the other quantities.

As  suggested by the density profiles at equilibrium, the analyze of
the liquid flow will be separately  carried in two cylindrical domains:
\newline
$\quad - $ In the \emph{core}, containing the axis of the tube, where the liquid
density at equilibrium is independent of $r$,
\newline
$\quad - $ In the
\emph{boundary layer}, near the solid wall of the tube, where the
density gradient is significant. %The thickness of the boundary layer
% should be small but can be chosen arbitrary.
Based on the observations made in   Section III, the thickness
of the boundary layer is of the order of 4$\delta_l$.\\
Consequently, the equation of motion   is solved  in the two
different regions by using a small length
 parameter.  Using a matched asymptotic expansion,
 different analytic solutions are obtained in both zones. An
immediate outcome should be that the  inner part of the boundary layer
solution matches the outer part of bulk flow.

%%%%%%%%%%%%%%%%%%%%%%%%%%%%%%%%%%%%
\subsection{Liquid flow in the core}
%%%%%%%%%%%%%%%%%%%%%%%%%%%%%%%%%%%%
Due to $\epsilon \ll 1$, the main term of Eq. (\ref{massd}) yields
\begin{equation*}
(\tilde{r} \tilde{\rho} \tilde{u})_{\tilde{r}}= 0
\end{equation*}
and consequently,
\begin{equation*}
 \tilde{u} = \frac{\psi (\tilde{z})}{\tilde{r}\,\tilde{\rho}},
\end{equation*}
where $\psi$ is a function of $\tilde{z}$ only.
Since $\tilde{u}$ must be bounded when $\tilde{r}$ goes to zero, we get $\psi(\tilde{z}) = 0$ and consequently $\tilde{u}= 0$.\\
 Considering that $u(r,z) \equiv 0$, Eq. (\ref{massd}) yields
\begin{equation*} \label{massc}
(\tilde{\rho} \tilde{w})_{\tilde{z}}=0.
\end{equation*}
Then, the momentum equations become
\begin{eqnarray}
 0 &=& -\frac{Re}{M^2} \, \tilde{\rho}_{\tilde{r}} +\frac{1}{3}\,\epsilon\tilde{w}_{\tilde{r}\tilde{z}}
  + \frac{Re}{M^2}\, \tau^2 \tilde{\rho} \left[ \frac{1}{\tilde{r}} \left(\tilde{r} \tilde{\rho}_{\tilde{r}} \right)_{\tilde{r}}+\epsilon^2 \tilde{\rho}_{\tilde{z}\tilde{z}} \right]_{\tilde{r}},\label{moment1core}\\
\epsilon Re\, \tilde{\rho} \left(\tilde{w} \tilde{w}_{\tilde{z}} \right) &=& -\frac{Re}{M^2}\, \epsilon \tilde{\rho}_{\tilde{z}}+\frac{1}{\tilde{r}}\left(\tilde{r} \tilde{w}_{\tilde{r}}\right)_{\tilde{r}}+\frac{4}{3}\, \epsilon^2 \tilde{w}_{\tilde{z}\tilde{z}}
 + \frac{Re}{M^2}\, \epsilon \tau^2 \tilde{\rho}\left[ \frac{1}{\tilde{r}} \left(\tilde{r} \tilde{\rho}_{\tilde{r}} \right)_{\tilde{r}}+\epsilon^2 \tilde{\rho}_{\tilde{z}\tilde{z}} \right]_{\tilde{z}}.\qquad\qquad\label{moment2core}
  \end{eqnarray}

In agreement with the coefficient values of Table \ref{TableKey1}, the
main parts of the momentum equations are obtained by retaining the
dominant terms in Eqs. (\ref{moment1core}--\ref{moment2core}) :
\begin{equation} \label{momentumd}
\tilde{\rho}_{\tilde{r}}=0  \qquad {\rm and}  \qquad \frac{1}{\tilde{r}}\left(\tilde{r} \tilde{w}_{\tilde{r}}\right)_{\tilde{r}}=\frac{Re}{M^2}\, \epsilon\, \tilde{\rho}_{\tilde{z}}.
\end{equation}
Note that, due to  $\tilde{\rho}_{\tilde{r}}=0$, the term $ \displaystyle \frac{Re}{M^2}\, \epsilon \tau^2
\tilde{\rho}\frac{1}{\tilde{r}} \left(\tilde{r}
  \tilde{\rho}_{\tilde{r}} \right)_{\tilde{r}\tilde{z}}$, which should
appear in the second equation (\ref{momentumd}) is null.
 % In the core we have $\epsilon\, , \tau \ll 1$ and therefore we obtain as

Equations (\ref{momentumd}) can be explicitly integrated and yield
 \begin{equation}
\tilde{\rho}=\tilde{\rho}(\tilde{z}) \qquad {\rm and}  \qquad \tilde{w}(\tilde{r},\tilde{z})=-\frac{Re}{4\, M^2}\, \epsilon \tilde{\rho}' \left(k_0-\tilde{r}^2\right)\label{rhovc}
\end{equation}
where $k_0$ is a constant to be determined by the boundary conditions.
Introducing this velocity field in the continuity equation we obtain
$$
\tilde{\rho}(\tilde{z})=\sqrt{h_0 \tilde{z}+h_1},
$$
where the constants $h_0$ and $h_1$ must be determined from the
inlet  and outlet bulk densities. For example, if we assume that
the inlet bulk density is $\tilde{\rho}(0)=1$, the outlet bulk density
$\tilde{\rho}(1)$ derives from Eq. (\ref{presure})  when
  $\func{grad}\Delta P= -10^6$ :
\[
\rho(L)-\rho(0)=\frac{\Delta P}{c_l^2}\simeq -0.46 \times 10^{-6}.
\]
Consequently, $h_1=1$ and $h_0=-0.92\times 10^{-6}$.
%%%%%%%%%%%%%%%%%%%%%%%%%%%%%%%%%%%%
\subsection{Liquid flow in the boundary layer}
%%%%%%%%%%%%%%%%%%%%%%%%%%%%%%%%%%%%
In the boundary layer, $\tilde{r}$ is always different from zero and
the reasoning made in  Section IV no longer works. From
Eq. (\ref{massd}) we get that $ \tilde{u}$ is of order of
$\epsilon\tilde{w}$. Then, introducing $\bar{u}$ as
\begin{equation*}
\tilde{u} = \epsilon \bar{u},
\end{equation*}
the continuity equation becomes
\begin{equation*}
\frac{1}{\tilde{r}}(\tilde{r} \tilde{\rho} \bar{u})_{\tilde{r}}+ (\tilde{\rho} \tilde{w})_{\tilde{z}}=0.\label{massd1}
\end{equation*}

To have an idea of what happens near the wall of the nanotube, we
have to translate and re-scale $\tilde{r}$ such that $ \tilde{r}=1/2-\xi
\overline{r}$. Hence, on the boundary of the nanotube where
$\tilde{r}=1/2$, we get $\overline{r}=0$. The value of $\xi$ is
determined by the condition  $\overline{r}=1$ on the separating
surface between the core and the boundary layer where
$\tilde{r}=1/2-4\tau$, and we get $\xi=4\tau$. Therefore, the
continuity equation is :
\begin{equation*}
  \label{massbl}
-\frac{1}{1- 2 \xi \overline{r}}\,[(1-2 \xi \overline{r}) \tilde{\rho}
\bar{u}]_{\overline{r}}+   \xi (\tilde{\rho} \tilde{w})_{\tilde{z}}=0,
\end{equation*}
and the momentum equations are :
\begin{eqnarray}
Re \,\epsilon^2 \tilde{\rho} \left(-\xi \bar{u}\bar{u}_{\overline{r}}
+ \xi^2  \tilde{w} \bar{u}_{\tilde{z}}\right)
&=&
\frac{Re}{M^2} \, \xi \tilde{\rho}_{\overline{r}}+
\frac{4}{3}\,\epsilon \left[\frac{1}{1-2 \xi\overline{r}}\left((1-2\xi
    \overline{r})\bar{u}\right)_{\overline{r}}\right]_{\overline{r}}
\nonumber\\\
& &\hspace*{-4cm}+\epsilon^3 \xi^2 \bar{u}_{\tilde{z}\tilde{z}}
-\frac{1}{3}\, \epsilon\, \xi \tilde{w}_{\overline{r}\tilde{z}}
- \frac{Re}{16\,M^2}\,  {\xi\, \tilde{\rho}}  \left[ \frac{1}{1-2\xi \overline {r}} \left((1-2\xi \overline{r}) \tilde{\rho}_{\overline{r}} \right)_{\overline{r}}+\epsilon^2 \xi^2\tilde{\rho}_{\tilde{z}\tilde{z}} \right]_{\overline{r}},\label{ml1}  \\
Re\, \epsilon \,\tilde{\rho} \left(-\xi\bar{u}\tilde{w}_{\overline{r}}
+ \xi^2 \tilde{w} \tilde{w}_{\tilde{z}}\right)
&=&
-\frac{Re}{M^2}\, \epsilon\, \xi^2
\tilde{\rho}_{\tilde{z}}+\frac{1}{1-2\xi \overline{r}}\left[ (1-2\xi
  \overline{r}) \tilde{w}_{\overline{r}} \right]_{\overline{r}}
+\frac{4}{3}\, \epsilon^2 \xi^2  \tilde{w}_{\tilde{z}\tilde{z}}\nonumber\\
& &\hspace*{-4cm}-\frac{1}{3}\, \epsilon^2 \xi\left[ \frac{1}{1-2\xi
    \overline{r}} \left((1-2\xi \overline{r}) \bar{u}
  \right)_{\overline{r}}\right]_{\tilde{z}}
+ \frac{Re}{16\,M^2}\, \epsilon\, \xi^2 \tilde{\rho} \left[ \frac{1}{1-2\xi \overline {r}} \left((1-2\xi \overline{r}) \tilde{\rho}_{\overline{r}} \right)_{\overline{r}}+\epsilon^2\xi^2 \tilde{\rho}_{\tilde{z}\tilde{z}} \right]_{\tilde{z}}.\qquad \label{ml2}
\end{eqnarray}

Then, neglecting the terms whose coefficients are very small, we
obtain from Eq. (\ref{ml1}) :
\begin{align}
 \label{momentub} \notag
& \frac{Re}{M^2}\,  \xi\, \tilde{\rho}_{\overline{r}}- \frac{Re}{16\,M^2} {\xi\,\tilde{\rho}}  \left[ \frac{1}{1-2\xi \overline {r}} \left((1-2\xi \overline{r}) \tilde{\rho}_{\overline{r}} \right)_{\overline{r}}  \right]_{\overline{r}}=0.
% \\  \notag
% & \underbrace{-\frac{Re}{M^2}\, \epsilon\, \xi^2\,
% \tilde{\rho}_{\tilde{z}}+
% \frac{Re}{16\,M^2}\, \epsilon\, \xi^2 \tilde{\rho}\, \frac{1}{1-2\xi
%   \overline {r}} \left[(1-2\xi \overline{r})
%   \tilde{\rho}_{\overline{r}}
% \right]_{\overline{r}\tilde{z}}}_\textrm{Je crois qu'on doit retenir
% aussi ces termes}
% +\frac{1}{1-2\xi \overline{r}}\left((1-2\xi\overline{r}) \tilde{w}_{\overline{r}}\right)_{\overline{r}}=0. \notag
\end{align}

This equation can be partially integrated and gives :
\[
\log(\tilde{\rho})-\frac{1}{16}  \left[ \frac{1}{1-2\xi \overline {r}} \left((1-2\xi \overline{r}) \tilde{\rho}_{\overline{r}} \right)_{\overline{r}}  \right]=k(\tilde{z})
\]
where $k$ is an unknown fonction of $\tilde{z}$ only. Then :
\begin{equation}
\tilde{\rho}_{\tilde{z}}-\frac{1}{16} \tilde{\rho} \left[
  \frac{1}{1-2\xi \overline {r}} \left((1-2\xi \overline{r})
    \tilde{\rho}_{\overline{r}} \right)_{\overline{r}}
\right]_{\tilde{z}}=\tilde{\rho} k^\prime(\tilde{z}).\label{eqkprime}
\end{equation}
Taking account of Eq. (\ref{eqkprime}), the dominant terms of
Eq. (\ref{ml2}) write :
 \[
-\frac{Re}{M^2}\, \epsilon\, \xi^2\,
\tilde{\rho}_{\tilde{z}}+
\frac{Re}{16\,M^2}\, \epsilon\, \xi^2 \tilde{\rho}\, \frac{1}{1-2\xi
  \overline {r}} \left[(1-2\xi \overline{r})
  \tilde{\rho}_{\overline{r}}
\right]_{\overline{r}\tilde{z}}=-\frac{Re}{M^2}\, \epsilon\, \xi^2\,\tilde{\rho} k^\prime(\tilde{z})
 \]
which should be equal to zero; therefore $k(\tilde{z})$ is constant
and Eq. (\ref{ml2}) is restricted to :
\[
\frac{1}{1-2\xi \overline{r}}\left((1-2\xi\overline{r})
  \tilde{w}_{\overline{r}}\right)_{\overline{r}}=0.
\]

The solution of this equation with the no-slip boundary condition is
\begin{equation}
\tilde{w}=-\frac{m(\tilde{z})}{2 \xi} \log(1-2\xi
\overline{r})
=-\frac{m(\tilde{z})}{2 \xi} \log(2\tilde{r})\label{vbl},
\end{equation}
where $m(\tilde{z})$ is a function to be determined with the continuity
condition of the velocity field through the surface separating the core
and the boundary layer, {\it i.e.} for $\tilde{r}=1/2-\xi$. From
Eqs. (\ref{rhovc}) and (\ref{vbl}) we get :
\begin{equation*}
-\frac{Re}{4 M^2}\, \epsilon \tilde{\rho}^\prime (\tilde{z})\,
\left[k_0-\left(\frac{1}{2}-\xi\right)^2\right]
=-\frac{m(\tilde{z})}{2  \xi} \log(1-2\xi).
\end{equation*}
Therefore $m$ is proportional with $\tilde{\rho}^\prime$ :
\begin{equation}
  \label{eqm}
  m(\tilde{z})=\frac{\xi\,\epsilon Re}{8 M^2}
  \frac{4 k_0-(1-2\xi)^2}{ \log(1-2\xi)}\,\tilde{\rho}^\prime (\tilde{z})
\end{equation}

\subsection{Velocity profile in the nanotube}

From Eqs. (\ref{rhovc}) (\ref{vbl}) and (\ref{eqm}) the expression of
the velocity field $\tilde{w}(\tilde{r},\tilde{z})$ in the whole domain
is :
\begin{equation}
  \label{velocity}
  \tilde{w}(\tilde{r},\tilde{z})=\left\{
\begin{array}{l}\displaystyle
-\frac{\epsilon Re}{4\, M^2}\, \tilde{\rho}'(\tilde{z})
\left(k_0-\tilde{r}^2\right),\qquad 0\leqslant \tilde{r}\leqslant
\frac{1}{2}-\xi\\[0.3cm]
\displaystyle-\frac{\epsilon Re}{16 M^2}
  \frac{4 k_0-(1-2\xi)^2}{ \log(1-2\xi)}\,\tilde{\rho}^\prime
  (\tilde{z})
\log(2\tilde{r}),\qquad \frac{1}{2}-\xi\leqslant \tilde{r}\leqslant
\frac{1}{2}
\end{array}.
\right.
\end{equation}
It depends on the constant $k_0$ which is determined by the
following average condition (which expresses the fact that the average
of $w$ on the outlet section of the tube is equal to $\hat{w}$) :
\[
\frac{4}{\pi}\int_0^{2\pi}\int_0^{1/2}\tilde{w}(\tilde{r},1)\,\tilde{r}\,
\textrm{d}\tilde{r}\textrm{d}\theta=1\quad\Leftrightarrow\quad
\int_0^{1/2}\tilde{w}(\tilde{r},1)\,\tilde{r}\,
\textrm{d}\tilde{r}=\frac{1}{8}.
\]
We obtain :
\begin{equation*}
  \label{kzero}
  k_0=\left(\frac{1}{2}-\xi \right)^2+
\frac{8+\alpha  (1-2 \xi )^4}{16 \alpha
   (1-\xi ) \xi } \log (1-2 \xi ),
\end{equation*}
where $\alpha=\displaystyle\frac{\epsilon Re}{4\, M^2}\,
\tilde{\rho}'(1)$ has a numerical value independent of the diameter of
the nanotube, $\alpha\simeq -8.0$.

\begin{figure}
  \begin{center}
    \includegraphics[width=6.4cm]{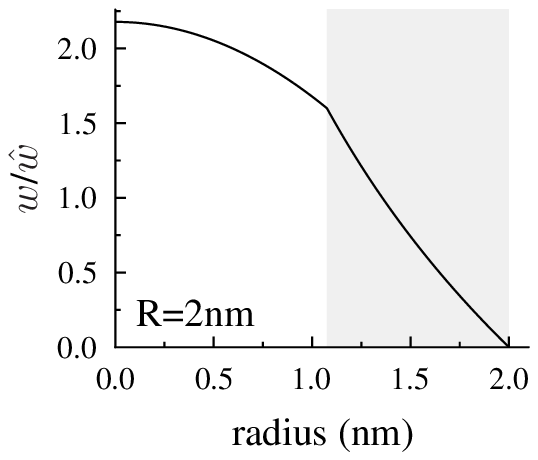}\qquad
    \includegraphics[width=6.4cm]{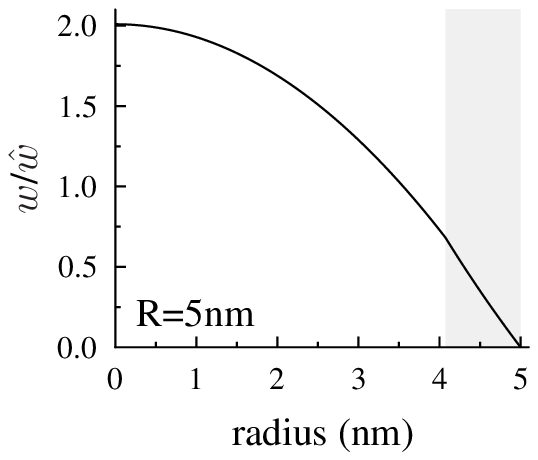}\\[0.1cm]
a)\hskip 7.2cm b)\\[0.3cm]
    \includegraphics[width=6.4cm]{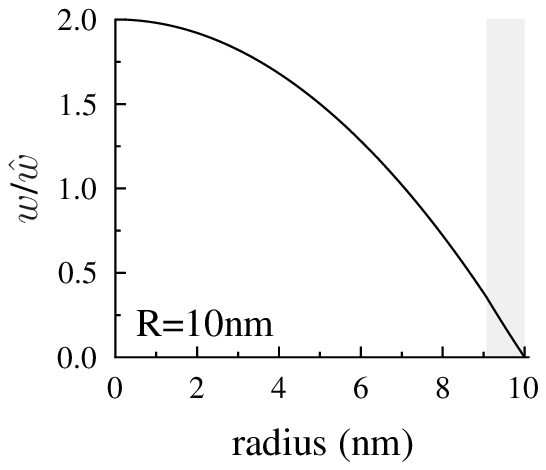}\qquad
    \includegraphics[width=6.4cm]{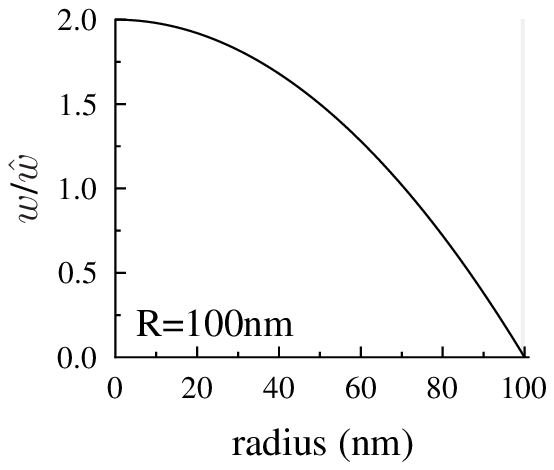}\\[0.1cm]
c)\hskip 7.2cm d)
  \end{center}
  \caption{Velocity profiles in nanotubes of different radius:
    (a) $R=2$ nm, (b) $R=5$ nm, (c) $R=10$ nm, (d) $R=100$ nm. The plots   simply show  the inner and outer regions up to and including a distinct point of separation of the two analytic representations. The region near the wall is represented in a grey area}
    \label{fig3}
\end{figure}

In Figure \ref{fig3} are plotted the profiles of the normalized
velocity $\tilde{w}$ (\ref{velocity}) in the four nanotubes.  The
motions are rather slow, the maximum of the velocity being about two
times $\hat{w}$ (see Table \ref{TableKey1}). As already mentioned, it is
assumed that the boundary layer (in grey on the Figure), where the
liquid is inhomogeneous, is the same than at equilibrium (see Section
3 and Fig. \ref{fig3}). Obviously, due to the condition (\ref{eqm}) the
graphs are continuous between the boundary layer and the core (see
Fig. \ref{fig3}).

\begin{figure}[t]
  \begin{center}
    \includegraphics{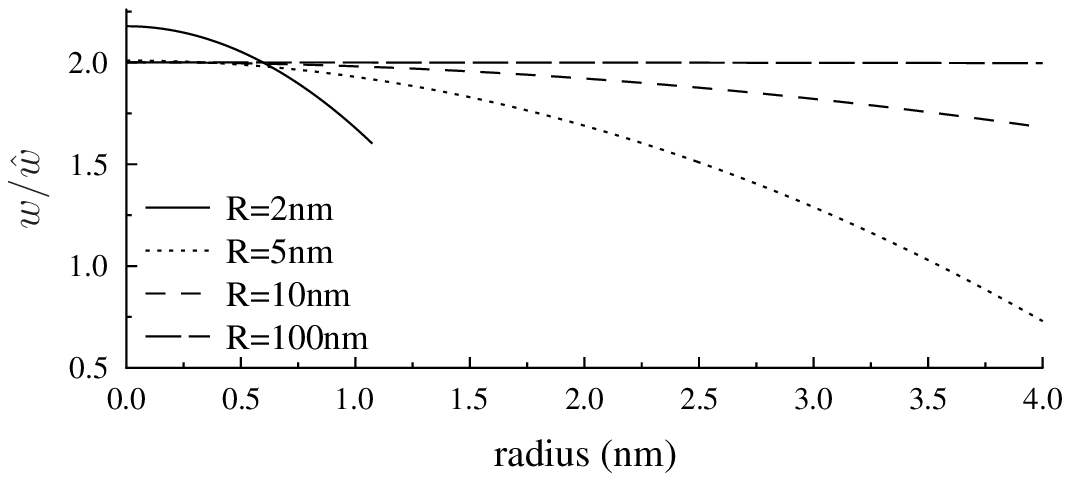}
  \end{center}
    \caption{Velocity profiles near the axis of the tube for four different nanotubes}
    \label{fig4}
\end{figure}

As for the classical Poiseuille flow, in the core the velocities
profiles are parabolic (see Eq. (\ref{rhovc})). In Figure \ref{fig4}
are plotted the profiles of the normalized velocity $\tilde{w}$ near
the axis of the tube, in the four nanotubes. For larger nanotubes (5 nm
to 100 nm) the normalized velocity is almost the same and the influence
of the boundary wall on the normalized velocity in the core is less
important that in the case of a thin tube (2 nm). It is worth noting
that the flow near the axis of a thin nanotube is ``proportionally''
faster that the flow in larger nanotubes.

Since the function $\tilde{\rho}^\prime(\tilde{z})$ has a weak
variation inside the interval $[0,1]$, the value of the velocity
(\ref{velocity}) at the interface between the core and the boundary
layer can be approximated by :
\[
\left.\tilde{w}\right|_{\tilde{r}=1/2-\xi}=-\frac{\tilde{\rho}^\prime(\tilde{z})}{\tilde{\rho}^\prime(1)}\frac{8+\alpha  (1-2 \xi
  )^4}{16 (1-\xi ) \xi } \log (1-2 \xi )\simeq-\frac{8+\alpha  (1-2 \xi
  )^4}{16 (1-\xi ) \xi } \log (1-2 \xi ).
\]

% \[
% \left.\frac{\partial\tilde{w}}{\partial r}\right|_{r=d/2}=
% -\frac{\tilde{\rho}^\prime(\tilde{z})}{\tilde{\rho}^\prime(1)}
% \frac{8+\alpha  (1-2 \xi
%   )^4}{8 (1-\xi ) \xi d} \simeq -\frac{8+\alpha  (1-2 \xi
%   )^4}{8 (1-\xi ) \xi d}
% \]

Whatever the radius of the nanotube, the density variation takes place
in a thin layer for a thickness of one nanometer. Outside of this
thin boundary layer, the liquid density is constant. In fact, due to
the very thin boundary layer, we may consider the motion as the motion
of an incompressible liquid in the core when $r \in [0, R-4\delta_l]$
and define a boundary slip velocity as the velocity obtained for $r =
R-4\delta_l$ corresponding to the frontier of the inhomogeneous liquid
layer. In this case, the Navier Length $b$
corresponds to  \cite{De Gennes}
\begin{equation}
\frac{w}{b}= \frac{\partial w}{\partial r} \qquad  {\rm when}\qquad r = R-4\delta_l. \label{NavierLength}
\end{equation}
Due to $w=0$ when $r=R$, and the fact that the variations of $w$ in
the boundary layer are smooth enough, the graph of velocity in the
boundary layer is near a straight line. Then, for water at
$20^\texttt{o}$ Celsius, the Navier length $b$ corresponds to the
boundary layer thickness which is about one nanometer and, due to
Eq. (\ref{meanw}), the slip velocity is
\begin{equation}
w_g= w_{_{|r =R-4\delta_l}} =  \tilde{w}_{_{|\tilde{r}=1/2-\xi}}\,\hat{w} =  \frac{8+\alpha  (1-2 \xi
  )^4}{16 (1-\xi ) \xi } \log (1-2 \xi ) \frac{d^2\,\func{grad} \Delta P}{32\, \kappa}.\label{fluid_velocity}
\end{equation}
Consequently, Eqs. (\ref{NavierLength}) and (\ref{fluid_velocity})
yield the boundary conditions for the Hagen-Poiseuille flow in the
core.

The values of the slip velocity $w_g$ are given in Table  \ref{TableKey2}
(in {\bf c.g.s.} system units). The case
when $R=100$ nm is close from a flat thin boundary layer and in our
model, the Navier length is constant whatever  the radius of the
nanotube is.

\begin{table}[tbp]
\centering
\begin{tabular}{|c||c|c|c|c|}
\hline\hline
d\,=\,2\,R & $4\times 10^{-7}$ & $10^{-6}$ & $2\times 10^{-6}$ &  $2\times 10^{-5}$  \\
\hline
$w_g$& $8.0\times 10^{-7}$ & $2.127\times 10^{-6}$ & $4.424\times
10^{-6}$ &  $4.603\times10^{-5}$   \\
\hline\hline
\end{tabular}
\vskip 0.15cm
\caption{Numerical values of the slip velocity $w_g$ following the radius value of the nanotube}
\label{TableKey2}
\end{table}

\section{Conclusion and comments}
The question about the \emph{correct} set of boundary conditions at
the nanoscale is recurrent in both molecular dynamics simulation and
the applications of continuum fluid-mechanics. Clearly, the classical
no-slip boundary condition of macroscopic fluid mechanics does not
apply, and in confined nano-flows, it is necessary to get a deep understanding
of the interfacial friction phenomena between   fluid and   wall.

Using the classical terminology we say that the slip velocity is the
tangential velocity of the fluid at the solid wall determined by a
surface friction coefficient $k$, while the Navier length represents
the length given by the ratio $k/\eta$ (\cite{De Gennes},
Fig. 1). Here we use a continuum model generalizing Navier-Stokes
equation via an internal energy function of the deformation and the
surdeformation of the fluid. For this reason the boundary effects
predicted by the model are deeply different from what we see in
classical Navier-Stokes equations.  This model accounts for an
embedding effect at the solid surfaces where the liquid is subjected
to strong variations of density. The intermolecular forces, mainly by
capillarity effects, create an inhomogeneous layer at the wall where
slippage of the liquid is possible. The thickness of the layer depends
on the molecular length $\delta_l$ and consequently on the temperature
through the surdeformation coefficient $\lambda$ of the fluid and the
isothermal sound speed ${c}_l$. The results are compatible with MD
simulations: the Gibbs adsoption is of the same order and the
inhomogeneous density layer has the same thickness in the two models.
The thickness of the inhomogeneous layer is the Navier length; the
slip velocity is the fluid velocity evaluated at the internal boundary
of the inhomogeneous layer.

Finally, the simple proposed model highlights the following points:\newline
The continuum mechanics approach is in intuitive agreement with what is expected by experiments and confirms the adequation of van der Waals' model in nanoscale framework by using a convenient representation of the fluid-solid interaction. \newline
The continuum mechanics approach is important to obtain simple analytical solutions for simple flow geometries.
\newline

\textbf{Acknowledgements}:
GS is partially supported by PRIN project 'Matematica e meccanica dei sistemi biologici e dei tessuti molli'; GS \& HG are also supported by 'Institut Carnot Star' for the stays during the year 2012 and the collaboration between Aix-Marseille Universit\'e and Universit\`{a} degli Studi di Perugia.

\end{document}